\documentclass[reprint,superscriptaddress,amsmath,amssymb,aps,prl,floatfix,twocolumn]{revtex4-2}

\usepackage{graphicx}
\usepackage{dcolumn}
\usepackage{bm}
\usepackage{hyperref}

\usepackage{amsmath}
\usepackage{amsfonts}
\usepackage{amssymb}
\usepackage{amsmath}
\usepackage{mathtools}
\usepackage{color}

\usepackage[utf8]{inputenc}
\usepackage[T1]{fontenc}

\usepackage{graphics}
\usepackage{graphicx}

\usepackage{amsfonts}

\usepackage{mathptmx}
\DeclareMathAlphabet{\mathcal}{OMS}{cmsy}{m}{n}
\usepackage{times}

\begin{document}

\title{Multidimensional Attosecond Clocking near Dirac Cones in Graphite}

\author{Daniel Kroeger}
\affiliation{Solid State Institute, Technion---Israel Institute of Technology, Haifa 32000, Israel}
\affiliation{The Helen Diller Quantum Center, Technion---Israel Institute of Technology, Haifa 32000, Israel}
\affiliation{The Norman Seiden Multidisciplinary Graduate Program in Nanoscience and Nanotechnology, Technion---Israel Institute of Technology, Haifa 32000, Israel}

\author{Anna Hassine}
\affiliation{Department of Physics, Technion---Israel Institute of Technology, Haifa 32000, Israel}
\affiliation{Solid State Institute, Technion---Israel Institute of Technology, Haifa 32000, Israel}
\affiliation{The Helen Diller Quantum Center, Technion---Israel Institute of Technology, Haifa 32000, Israel}

\author{Eyal Uzner}
\affiliation{Schulich Faculty of Chemistry, Technion---Israel Institute of Technology, Haifa 3200003, Israel}
\affiliation{Department of Physics, Technion---Israel Institute of Technology, Haifa 32000, Israel}

\author{Camilo Granados}
\affiliation{Eastern Institute of Technology, Ningbo 315200, China}

\author{Marcelo F. Ciappina}
\affiliation{Department of Physics, Technion---Israel Institute of Technology, Haifa 32000, Israel}
\affiliation{Department of Physics, Guangdong Technion---Israel Institute of Technology, Shantou 515063, Guangdong, China}
\affiliation{Guangdong Provincial Key Laboratory of Materials and Technologies for Energy Conversion, Guangdong Technion---Israel Institute of Technology, Shantou 515063, Guangdong, China}

\author{Ofer Neufeld}
\altaffiliation{Corresponding author: ofern@technion.ac.il}
\affiliation{Schulich Faculty of Chemistry, Technion---Israel Institute of Technology, Haifa 32000, Israel}

\author{Michael Kr\"uger}
\altaffiliation{Corresponding author: krueger@technion.ac.il}
\affiliation{Department of Physics, Technion---Israel Institute of Technology, Haifa 32000, Israel}
\affiliation{Solid State Institute, Technion---Israel Institute of Technology, Haifa 32000, Israel}
\affiliation{The Helen Diller Quantum Center, Technion---Israel Institute of Technology, Haifa 32000, Israel}

\author{Zhaopin Chen}
\altaffiliation{Corresponding author: zhaopin.chen@campus.technion.ac.il}
\affiliation{Department of Physics, Technion---Israel Institute of Technology, Haifa 32000, Israel}
\affiliation{Solid State Institute, Technion---Israel Institute of Technology, Haifa 32000, Israel}
\affiliation{The Helen Diller Quantum Center, Technion---Israel Institute of Technology, Haifa 32000, Israel}

\date{\today}

\begin{abstract}
\noindent
Two-color high-harmonic spectroscopy is widely used to access sub-cycle electron dynamics and to retrieve harmonic timing information, including harmonic phases and attochirp across gases, solids, and liquids. However, the dependence of such timing observables on additional laser-control parameters remains largely unexplored. Here, we introduce driving intensity as an additional dimension of two-color harmonic spectroscopy in highly oriented pyrolytic graphite (HOPG). The retrieved attosecond two-color delays maximizing the 4th and 5th harmonic yields evolve systematically and differently with driving intensity. Semiconductor Bloch-equation calculations reproduce these trends and reveal a pronounced sensitivity of the intensity-dependent delays to the electronic band dispersion. Our results demonstrate multidimensional attosecond clocking near Dirac cones and establish intensity-dependent two-color delays as a sensitive observable for band-dispersion in quantum materials.


\end{abstract}

\maketitle



\noindent Strong laser fields can drive electrons in atomic gases \cite{McPherson1987,Ferray1988} and solids \cite{Ghimire2011,Ghimire2019} far beyond the perturbative regime, resulting in the emission of high-energy photons via high-harmonic generation (HHG). In atomic targets, HHG is well described by the three-step model consisting of tunnel ionization, field-driven acceleration, and recombination with the parent ion\cite{corkum1993plasma}. In solids, HHG arises from the dynamics of interband polarization and intraband currents\cite{Ghimire2019}. The interband contribution is associated with electron–hole pair creation and recombination (coherence), while the intraband contribution originates from laser-driven carrier motion within the bands. 

Over the past decade, solid-state HHG has developed into a powerful spectroscopic tool for probing ultrafast quasiparticle dynamics on sub-cycle timescales under strong-field driving \cite{Vampa2015a,Heide2024}. It provides access to attosecond electron–hole dynamics and enables the investigation of a wide range of condensed-matter phenomena, including band-structure retrieval and reconstruction~ \cite{vampa2016crystal,uzan2022observation}, Berry curvature and Berry phase effects ~\cite{borsch2023lightwave,uzan2024observation,bai2024probing}, topological surface states and phase transitions ~\cite{bai2021high,silva2019topological,heide2022probing,PhysRevX.13.031011}, and valley-selective dynamics ~\cite{jimenez2020lightwave,tyulnev2024,mitra2024light}. Furthermore, solid-state HHG provides a route toward compact, high-repetition-rate attosecond pulse sources in the extreme-ultraviolet (XUV) and deep-ultraviolet (DUV) regimes~\cite{Garg2016Nature,Nayak2024,chen2025attosecond}.

Dirac materials provide a particularly compelling setting for solid-state HHG. In graphene, the linear, gapless dispersion near the Dirac points enables the study of relativistic-like quasiparticles in a condensed-matter system \cite{novoselov2005two}. These distinctive electronic properties have made Dirac systems an important platform for exploring light-induced Floquet phenomena \cite{oka2009photovoltaic,mciver2020light,H.2013,lesko2026probing,Lesko2026a}, while intense laser driving further accesses their highly nonlinear optical response, including HHG ~\cite{ExoticGraph2,Hassan2018extremely}. In contrast to conventional semiconductors with a finite band gap and approximately parabolic dispersion near the band edges, the gapless linear dispersion in Dirac materials leads to strong coupling between interband polarization and intraband currents \cite{al2014high}. This interplay produces distinctive harmonic responses, including a nontrivial dependence on laser ellipticity~\cite{ExoticGraph2,sato2021high}, and femtosecond time shifts in the emission of nonperturbative harmonics around the Dirac cone~\cite{Chen2026}. 

The emitted harmonics and their spectrum carry the fingerprint of the internal attosecond electron-hole dynamics. However, only a multi-dimensional ``clocking'' measurement can reveal attosecond temporal information. Two-color high-harmonic spectroscopy, where the modulation of the harmonic yield is recorded as a function of delay between two fields of different colors, is a powerful approach to recover ultrafast dynamics in gases~\cite{NDudovich,Shafir2012,Pedatzur2015,Kneller2022} and solids~\cite{Vampa2015a,uzan2022observation,uzan2024observation}. The relative \(\omega\)-\(2\omega\) phase provides a calibrated sub-cycle temporal reference, and the phase maximizing a given harmonic yields defines the clocking observable. Its temporal precision is therefore set by the experimental resolution of the relative two-color phase. While its variation across harmonic orders is commonly used to retrieve the attochirp, i.e., the frequency-dependent emission time of the harmonics~\cite{NDudovich}, its dependence on other experimental control parameters remains largely unexplored. 

Here, we extend this clocking approach to the driving-intensity dimension and apply it to highly oriented pyrolytic graphite (HOPG), a bulk material composed of stacked graphene layers~\cite{Pappis1961,Blackman1962}. HOPG offers a damage-resilient alternative to monolayer graphene. Near the Dirac points, HOPG preserves the essential Dirac-like dispersion of graphene while providing superior mechanical stability and a substantially higher damage threshold~\cite{CarSynth,HOPG,zhou2006first}, making it well suited for strong-field and two-color HHG experiments.  At each pump intensity, we determine the relative two-color delay that maximizes the H4 or H5 yield and track shifts of this optimum with attosecond sensitivity. These maximizing delays evolve systematically with intensity and exhibit a weaker intensity dependence at the highest fields. Semiconductor Bloch equation (SBEs) calculations reproduce the measured trends and demonstrate their pronounced sensitivity to the long-range hopping terms and the resulting band curvature. Intensity-dependent two-color delays therefore provide a sensitive observable for constraining realistic electronic band dispersions.

\begin{figure}[htb!]
    \centering
    \includegraphics[width=1\columnwidth]{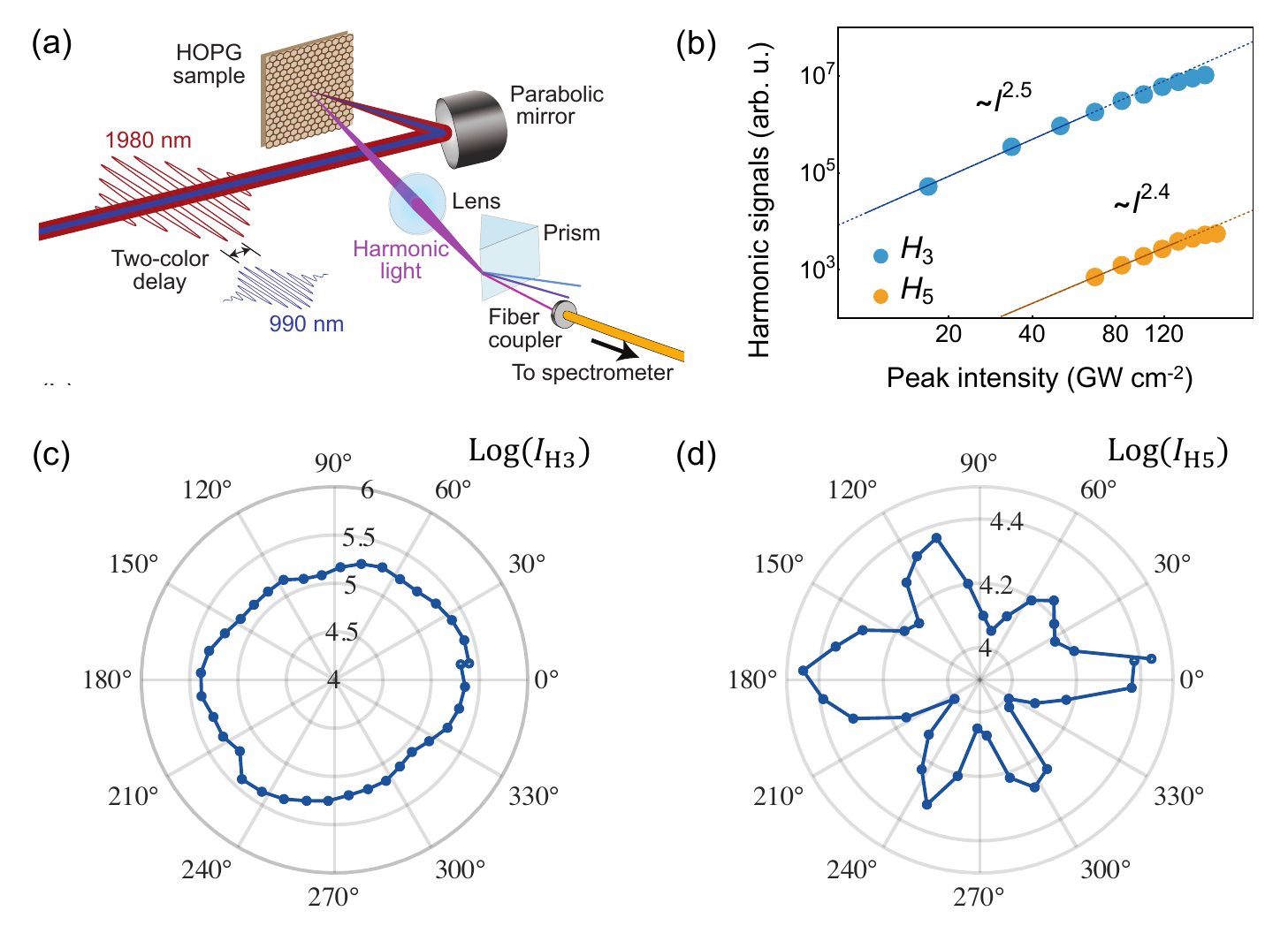}
    \caption{Non-perturbative harmonic generation from graphite. (a) Experimental setup for two-color spectroscopy. A two-color laser field comprised of a fundamental (red) and second-harmonic (blue) pulse is focused onto a sample of highly-ordered pyrolytic graphite (HOPG), producing harmonics in reflection. (b) Double-logarithmic plot of the intensity scaling of the harmonic signal of H3 (blue circles) and H5 (yellow circles). The lines correspond to power-law fit curves which is based on the data points where fit curve is displayed as a solid curve. (c,d) Pump-polarization dependence of the H3 and H5 yields, exhibiting nearly isotropic and sixfold patterns, respectively. }
    \label{fig1}
\end{figure}

In our experiment, the fundamental driving pulse (frequency $\omega_0$) has a central wavelength of 1980\,nm, a duration of 60\,fs, and a repetition rate of 1\,MHz. The pulse is generated by an optical parametric chirped pulse amplifier laser system (Class 5 Photonics White Dwarf~\cite{Braatz2021}). A time-delayed weak second-harmonic field (frequency $2\omega_0$), with an intensity of 0.3\% of the fundamental and with identical polarization, is added as a perturbation in our two-color spectroscopy setup. The $\omega_0$ and $2\omega_0$ beams are combined collinearly and spatially overlapped, while their relative delay $\tau$ is precisely controlled with attosecond resolution (see Fig.~\ref{fig1}(a) for a sketch of the experimental setup, and the Supplemental Material for more details). 

An off-axis parabolic mirror ($f=25$\,mm) tightly focuses the two-color pulses onto an HOPG sample (TipsNano), generating harmonics in reflection~\cite{Vampa2018}. A lens refocuses the reflected driving fields and generated harmonics, which are then spatially separated by a prism according to wavelength. This arrangement suppresses the intense fundamental and second-harmonic fields. The selected harmonic is collected by an optical fiber, and its spectrum is recorded using a Si-based spectrometer.

In absence of the weak $2\omega_0$ pulse, we observe the generation of the third and the fifth harmonic (H3 and H5). We find that the signal of both harmonics does not follow the perturbative scaling with intensity $I$, $\sim I^n$, where $n$ is the harmonic number (see Fig.~\ref{fig1}(b)). Instead, the scaling is non-perturbative, indicating that carrier acceleration and recombination dominate the harmonic generation rather than multiphoton transitions~\cite{ExoticGraph2,Avetissian2022}. While the harmonics observed in our HOPG experiment are not `high' harmonics, they still exhibit non-perturbative scaling which is a signature of the strong-field regime and HHG.

We find that the H5 yield strongly depends on the pump polarization, exhibiting a sixfold pattern, whereas H3 is nearly isotropic (as seen in other hexagonal systems as well~\cite{kim2025quantum}). The H3 photon energy corresponds primarily to the approximately linear and isotropic Dirac-cone regime (see the band structure of HOPG in the SM). By contrast, H5 reaches the curved, non-Dirac region of the band structure, where the dispersion is more anisotropic and therefore more sensitive to the crystal orientation. This distinction may also explain why a sixfold orientation dependence was not resolved in earlier graphene HHG experiments~\cite{ExoticGraph2,cha2022gate}, in which the detected harmonics predominantly probed the Dirac regime. Our 2D SBEs calculations further show that the H5 yield is maximized when the pump polarization is aligned along the $\Gamma$-M direction, corresponding to the C–C bond direction.

\begin{figure}[htb!]
    \centering
    \includegraphics[width=0.75\columnwidth]{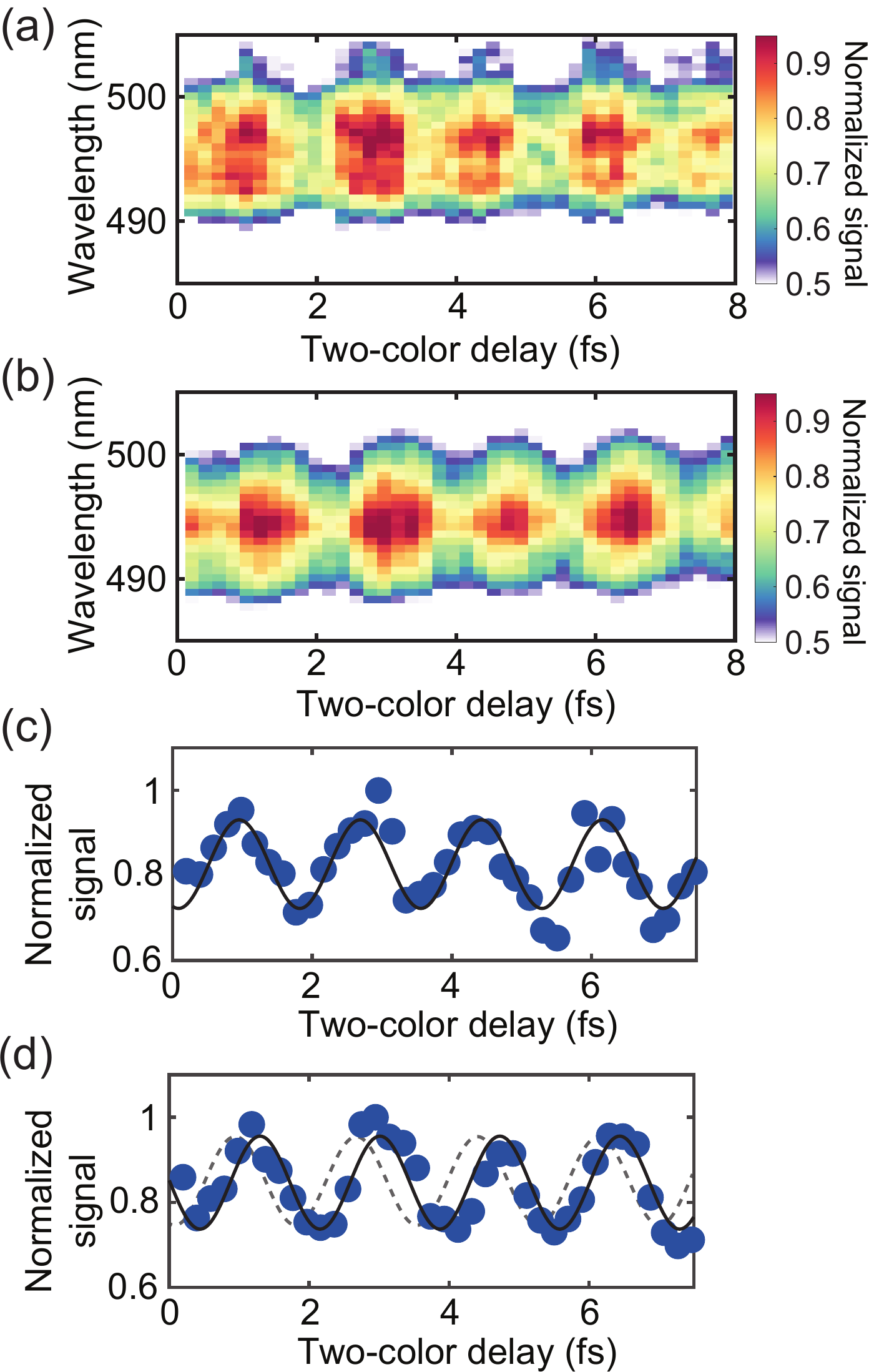}
    \caption{Sub-cycle two-color modulation of H4. (a,b) Measured spectra at two different pump intensities as a function of the two-color delay for $32\,\mathrm{GW\,cm}^{-2}$ and $48\,\mathrm{GW\,cm}^{-2}$, respectively. (c,d) Corresponding integrated H4 signals extracted from (a,b) along with sinusoidal fit curves. For comparison, we display the fit curve from (c) as a dashed line in (d). }
    \label{fig2}
\end{figure}

Adding the weak $2\omega_0$ field, we break the symmetry of the fundamental field and observe the generation of the even harmonic H4 in the $\Gamma$-M orientation. Scanning the two-color delay on sub-cycle time scales allows us to perform two-color spectroscopy with intrinsic attosecond resolution. We vary the laser intensity and study the intensity-dependent behavior of the two-color modulation while keeping the intensity ratio of the two colors fixed. Figure~\ref{fig2} shows the result of this measurement, comparing two intensities, $32\,\mathrm{GW\,cm}^{-2}$ (see Fig.~\ref{fig2}(a) and (c)) and $48\,\mathrm{GW\,cm}^{-2}$ (see Fig.~\ref{fig2}(b) and (d)). A subtle temporal shift of the modulation on the order of hundred attoseconds can be clearly seen (see Fig.~\ref{fig2}(d)).

\begin{figure}[htb!]
    \centering
    \includegraphics[width=0.8\columnwidth]{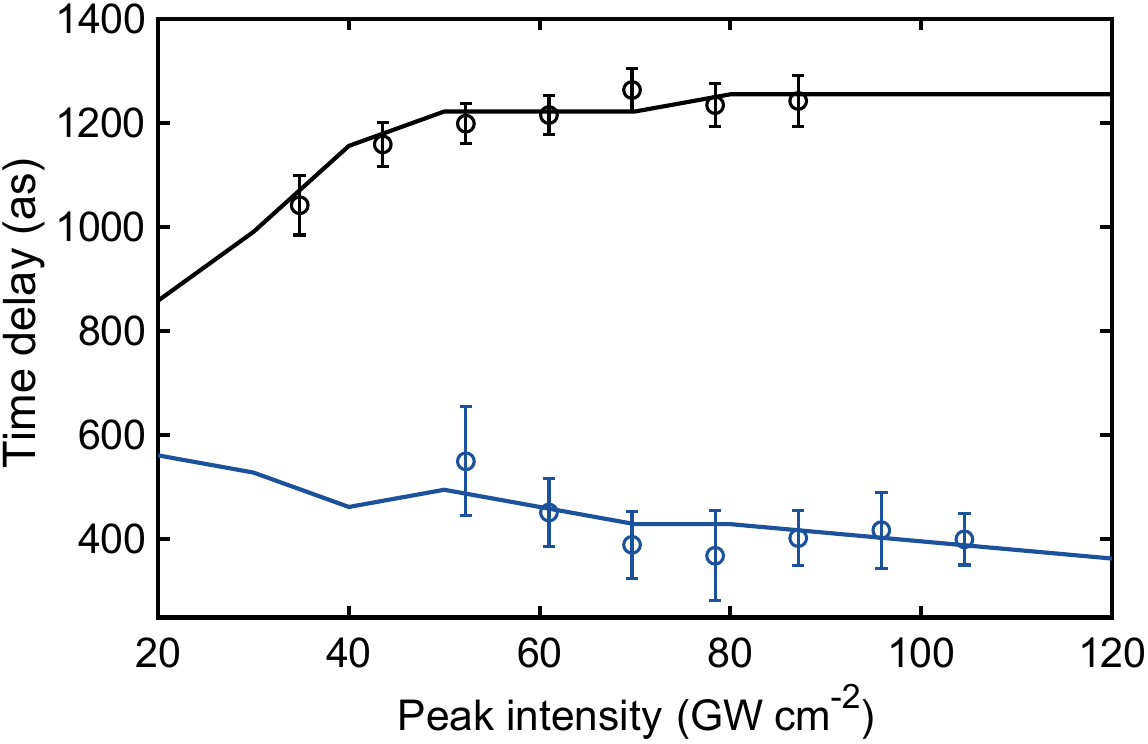}
    \caption{Intensity-dependent attosecond delays in HOPG. Measured two-color delays maximizing the H4 (black circles) and H5 (blue circles) yields as functions of peak intensity. Solid curves show semiconductor Bloch equation calculations. A constant vertical offset is applied to the experimental data to align them with the calculated delays. }
    \label{fig3}
\end{figure}

Next, we define the two-color delay that maximizes the H4 yield as the clocking delay, \(\tau_{\max}\). Repeating this clocking measurement independently at different driving intensities allows us to track the intensity dependence of \(\tau_{\max}\). Figure~\ref{fig3} is the central experimental result of this work. We observe that the attosecond time delay of H4 increases as a function of intensity in HOPG. Around $55\,\mathrm{GW\,cm}^{-2}$, the time delay shows a much weaker variation and a nearly flat behavior. 

We also perform two-color attosecond clocking on the odd harmonic H5 (black curve in Fig.~\ref{fig3}) which shows a strong modulation as a function of two-color delay. We find the opposite slope (decreasing attosecond time delay), followed by a flatter behavior starting at around $55\,\mathrm{GW\,cm}^{-2}$. Within HOPG, no further experimental ground for comparison exists because H6 and higher harmonics are not observed and H3 does not show any two-color modulation effect.

Figure~\ref{fig3} presents theoretical calculations based on the SBEs. In brief, we integrate the SBEs for monolayer graphene in two dimensions in the Houston basis using the experimental parameters and a dephasing time $T_2 = 5.7$\,fs (see SM and refs.~\cite{neufeld2026accurate,Herling2026} for further details). 2D SBEs calculations have yielded excellent agreement with much more complex graphite calculations in three dimensions~\cite{Chen2026}. We observe that experiment and theory agree well. For H4, the increase in time delay and the transition to a flat behavior are reproduced well. Also, a good agreement is found for H5. We note that the absolute attosecond timing on the vertical axis is not defined (only relative delays are measured) and the experimental data has been shifted vertically to match the theory curve (though the shift value is fixed for both H4 and H5, increasing our trust in the model).

\begin{figure}[htb!]
    \centering
 \includegraphics[width=1\columnwidth]{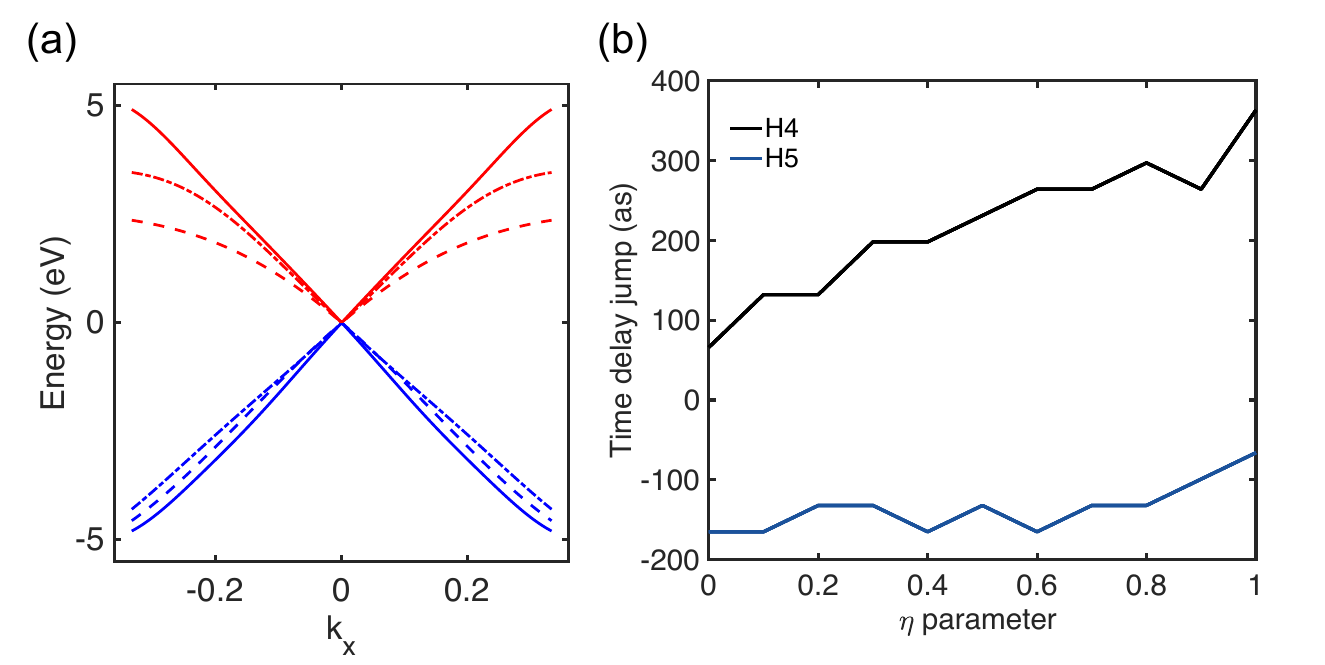}
    \caption{(a) Band dispersions near the K point along a momentum-space direction parallel to $\Gamma$–M. Dashed, dash-dotted, and solid curves denote the 2nd-, 5th-, and 14th-nearest-neighbor tight-binding models, respectively. (b) Calculated delay jump between $20$ and $50\,\mathrm{GW\,cm^{-2}}$ as a function of the long-range coupling parameter $\eta$. }
    \label{fig4}
\end{figure}

To elucidate the influence of band curvature on the attosecond delay, we define the delay jump as the difference between the calculated clocking delays, $\tau_{\mathrm{max} }$, at pump intensities of $20$ and $50\,\mathrm{GW\,cm^{-2}}$, which show the strongest effect. Our goal is to test if small changes to the band dispersion and Fermi velocity preserve this delay jump, or if it is ultra-sensitive to the bands. Figure~\ref{fig4}(a) compares the band dispersions near the K point parallel to the $\Gamma$–M direction obtained from three model levels: 2nd-, 5th-, and 14th-nearest-neighbor tight-binding models (see the SM for details). These models match at K exactly, but start differing away from K, where the 2nd NN model slightly underestimates the Fermi velocity. Figure~\ref{fig4}(b) shows the corresponding delay jumps as a function of the long-range coupling parameter $\eta$, which has been defined to smoothly interpolate between the 2nd-NN model at $\eta=0$ and the full 14th-NN model at $\eta=1$, which closely reproduces the DFT band structure \cite{neufeld2026accurate}. The pronounced variation of the delay jump with $\eta$ demonstrates its sensitivity to the detailed band curvature. The 14th-NN model yields the best agreement with the measured intensity dependence in Fig.~\ref{fig3}, while the H4 delay jump exhibits substantially greater sensitivity to band curvature than that of H5 (which can also be generated in non-$2\omega$ pathways). The 5th-NN model essentially reproduces the same delay jump, despite predicting unrealistic bands very far from K (not shown). Overall, these results suggest that intensity-dependent attosecond-delay measurements could provide a means of constraining realistic band-structure models and reconstructing the Fermi velocity, while variation in bands farther from K towards $\Gamma$ and M are ruled insignificant in our conditions (because the 14th NN and 5th-NN models differ greatly there, but still predict a similar phase jump).

\begin{figure}[htb!]
    \centering
 \includegraphics[width=0.82\columnwidth]{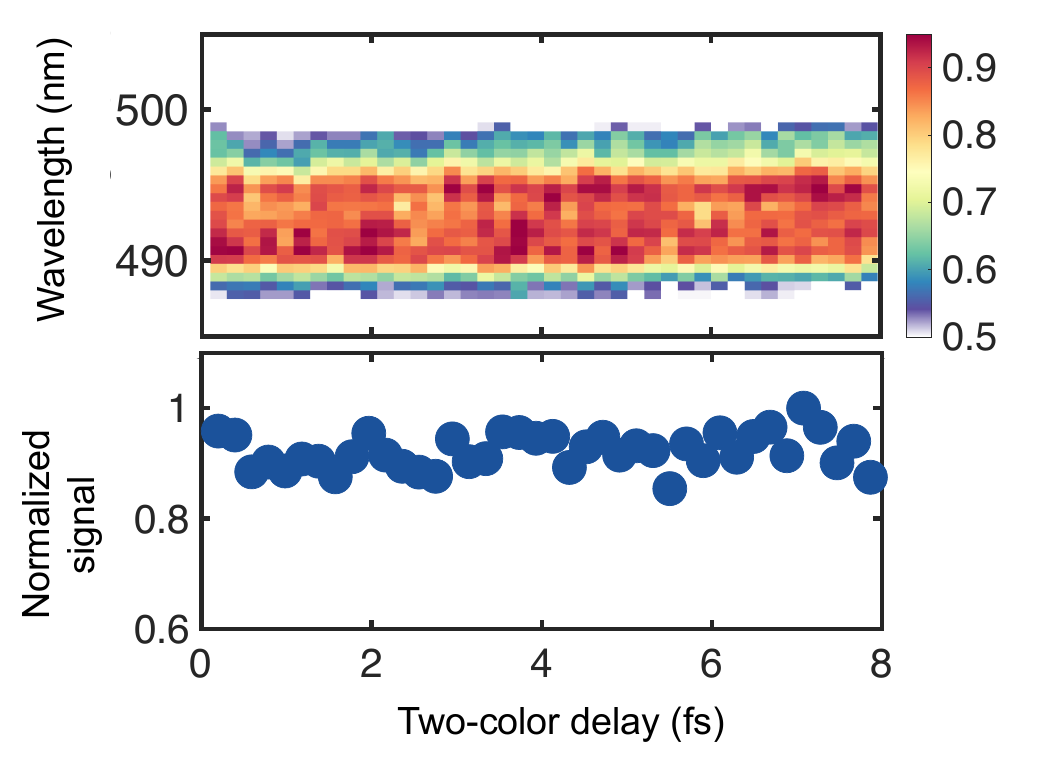}
    \caption{Measured H4 spectra (top panel) and integrated H4 yield (bottom panel) as functions of the two-color delay, recorded at the crystal orientation corresponding to the minimum H5 yield. In contrast to Fig.~\ref{fig2}, no obvious delay-dependent modulation is observed. }
    \label{fig5}
\end{figure}

To further assess the influence of crystal orientation on the two-color delay response, we measured the H4 yield as a function of delay at the orientation corresponding to the minimum H5 yield (see Fig.~\ref{fig5}). At this orientation, no discernible delay-dependent modulation is observed, preventing the reliable extraction of an attosecond delay. This result reveals a pronounced orientation dependence of the two-color response, which could be used to further expand multi-dimensional two-color HHG spectroscopy and should be addressed in a future work.

In conclusion, we demonstrate intensity-resolved two-color attosecond clocking measurements in HOPG, a Dirac material. The nearly isotropic H3 response and sixfold H5 pattern distinguish harmonic emission predominantly associated with the isotropic Dirac cone from that sampling higher optical non-linearities associated with interband transitions. The intensity-dependent shifts of the maximizing two-color phase provide an attosecond-sensitive clocking observable, and its systematic dependence on driving intensity is reproduced by SBE calculations and exhibits pronounced sensitivity to band curvature and Fermi velocity slightly away from the Dirac point. These complementary observables provide a potential route to test and reconstruct realistic electronic dispersions, complementary to attochirp or photocurrent approaches~\cite{vampa2016crystal,Weitz2024}, and might provide enhanced sensitivity. Future studies may resolve new electron–hole trajectories, advance subcycle lightwave electronics, and probe ultrafast decoherence mechanisms, including electron–phonon coupling.

\section*{Acknowledgements}
We thank M.~Ivanov and B.~Ma for insightful discussions and U.~Leonhardt and Y.~Rosenberg for providing specialized equipment. D.~K., C.~G., M.~F.~C., M.~K.~and Z.~C.~acknowledge the Guangdong Technion -- Israel Institute of Technology (GTIIT) and Technion Seed Grant Program for enabling their collaboration and joint research.
D.~K., O.~N., M.~K.~and Z.~C.~thank the Helen Diller Quantum Center and the Russell Berrie Nanotechnology Institute at the Technion for partial financial support. 
M.F.C. acknowledges support by the Quantum Science and Technology-National Science and Technology Major Project (Grant No. 2025ZD0301000),  the National Key Research and Development Program of China (Grant No. 2023YFA1407100), the Guangdong Province Science and Technology Major Project (Future functional materials under extreme conditions - 2021B0301030005) and the National Natural Science Foundation of China (Grant No. 12574092).


%

\end{document}